%% file: main.tex
\documentclass{article}

\usepackage[creativeai]{neurips_2025}
\usepackage{tipa}
\usepackage[utf8]{inputenc} 
\usepackage[T1]{fontenc}    
\usepackage{hyperref}       
\usepackage{url}            
\usepackage{booktabs}       
\usepackage{amsfonts}       
\usepackage{nicefrac}       
\usepackage{microtype}      
\usepackage{xcolor}         

\usepackage{graphicx}
\usepackage{subcaption}
\usepackage{natbib}
\title{8bit-GPT: Exploring Human-AI Interaction on Obsolete Macintosh Operating Systems}

\author{
 Hala Sheta \\
  Department of Computer Science\\
 University of Waterloo\\
  Waterloo, Canada\\
  \texttt{hsheta@uwaterloo.ca} \\
}

\begin{document}

\maketitle

\begin{abstract}
 The proliferation of assistive chatbots offering efficient, personalized communication has driven widespread over-reliance on them for decision-making, information-seeking and everyday tasks. This dependence was found to have adverse consequences on information retention as well as lead to superficial emotional attachment~\cite{spatola2024efficiency}. As such, this work introduces \texttt{8bit-GPT}; a language model simulated on a legacy Macintosh Operating System, to evoke reflection on the nature of Human-AI interaction and the consequences of anthropomorphic rhetoric. Drawing on reflective design principles such as slow-technology~\cite{hallnas2001slow} and counterfunctionality~\cite{pierce2014counterfunctional}, this work aims to foreground the presence of chatbots as a \textit{tool} by defamiliarizing the interface and prioritizing inefficient interaction, creating a friction between the familiar and not.
\end{abstract}

\input{sections/1-introduction}

\input{sections/2-related-work}
\input{sections/3.01-method}

\input{sections/4.01-user-study}
\input{sections/5-conclusion}
\input{sections/6-acknowledgements}
\small
\bibliography{main}
\newpage
\input{sections/7-appendix}





\newpage
\section*{NeurIPS Paper Checklist}

\begin{enumerate}

\item {\bf Claims}
    \item[] Question: Do the main claims made in the abstract and introduction accurately reflect the paper's contributions and scope?
    \item[] Answer: \answerYes{} 
    \item[] Justification: The abstract and introduction clearly state the claims and main approach of the paper.

\item {\bf Limitations}
    \item[] Question: Does the paper discuss the limitations of the work performed by the authors?
    \item[] Answer: \answerYes{} 
    \item[] Justification: Limitations related to the implementation of the concept are discussed in Section \ref{sec:conclusion}.

\item {\bf Theory assumptions and proofs}
    \item[] Question: For each theoretical result, does the paper provide the full set of assumptions and a complete (and correct) proof?
    \item[] Answer: \answerNA{} 
    \item[] Justification: The paper does not include theoretical results.

    \item {\bf Experimental result reproducibility}
    \item[] Question: Does the paper fully disclose all the information needed to reproduce the main experimental results of the paper to the extent that it affects the main claims and/or conclusions of the paper (regardless of whether the code and data are provided or not)?
    \item[] Answer: \answerYes{} 
    \item[] Justification: All implementation details including instructions on starting up the program on the emulator are included in Section \ref{sec:pipeline}. The repository is also provided (with an accompanying demo video).

\item {\bf Open access to data and code}
    \item[] Question: Does the paper provide open access to the data and code, with sufficient instructions to faithfully reproduce the main experimental results, as described in supplemental material?
    \item[] Answer: \answerYes{} 
    \item[] Justification: The repository is linked in Section \ref{sec:pipeline}, which includes a \texttt{README} on how to start the program. 

\item {\bf Experimental setting/details}
    \item[] Question: Does the paper specify all the training and test details (e.g., data splits, hyperparameters, how they were chosen, type of optimizer, etc.) necessary to understand the results?
    \item[] Answer: \answerNA{} 
    \item[] Justification: The paper does not include such experiments.

\item {\bf Experiment statistical significance}
    \item[] Question: Does the paper report error bars suitably and correctly defined or other appropriate information about the statistical significance of the experiments?
    \item[] Answer: \answerYes{} 
    \item[] Justification: The paper includes error bars for the quantitative analysis of the SUS scores in Appendix \ref{app:sus}.

\item {\bf Experiments compute resources}
    \item[] Question: For each experiment, does the paper provide sufficient information on the computer resources (type of compute workers, memory, time of execution) needed to reproduce the experiments?
    \item[] Answer: \answerYes{} 
    \item[] Justification: This is mentioned in Section \ref{sec:pipeline} and the \texttt{vec-inf} library deals with the resource allocation depending on the requested model.
    
\item {\bf Code of ethics}
    \item[] Question: Does the research conducted in the paper conform, in every respect, with the NeurIPS Code of Ethics \url{https://neurips.cc/public/EthicsGuidelines}?
    \item[] Answer: \answerYes{} 
    \item[] Justification: The user study protocol has been ethics-approved under REB \#47727. Efforts have been made to preserve anonymity where necessary. 

\item {\bf Broader impacts}
    \item[] Question: Does the paper discuss both potential positive societal impacts and negative societal impacts of the work performed?
    \item[] Answer: \answerYes{} 
    \item[] Justification: In Section \ref{sec:discussion}, the broader impacts of anthropomorphism and Human-AI interaction are discussed through the lens of analysing the artwork interactions.
    
\item {\bf Safeguards}
    \item[] Question: Does the paper describe safeguards that have been put in place for responsible release of data or models that have a high risk for misuse (e.g., pretrained language models, image generators, or scraped datasets)?
    \item[] Answer: \answerNA{} 
    \item[] Justification: The paper poses no such risks.

\item {\bf Licenses for existing assets}
    \item[] Question: Are the creators or original owners of assets (e.g., code, data, models), used in the paper, properly credited and are the license and terms of use explicitly mentioned and properly respected?
    \item[] Answer: \answerYes{} 
    \item[] Justification: All asset creators include the creator of emulator programs and models used are properly credited in Section \ref{sec:method}.

\item {\bf New assets}
    \item[] Question: Are new assets introduced in the paper well documented and is the documentation provided alongside the assets?
    \item[] Answer: \answerNA{} 
    \item[] Justification: The paper does not release new assets.

\item {\bf Crowdsourcing and research with human subjects}
    \item[] Question: For crowdsourcing experiments and research with human subjects, does the paper include the full text of instructions given to participants and screenshots, if applicable, as well as details about compensation (if any)? 
    \item[] Answer: \answerNA{} 
    \item[] Justification: The paper does not involve crowdsourcing nor research with human subjects.

\item {\bf Institutional review board (IRB) approvals or equivalent for research with human subjects}
    \item[] Question: Does the paper describe potential risks incurred by study participants, whether such risks were disclosed to the subjects, and whether Institutional Review Board (IRB) approvals (or an equivalent approval/review based on the requirements of your country or institution) were obtained?
    \item[] Answer: \answerYes{} 
    \item[] Justification: The study contains no such risks and an REB approval was obtained as detailed in Section \ref{sec:discussion}.

\item {\bf Declaration of LLM usage}
    \item[] Question: Does the paper describe the usage of LLMs if it is an important, original, or non-standard component of the core methods in this research? Note that if the LLM is used only for writing, editing, or formatting purposes and does not impact the core methodology, scientific rigorousness, or originality of the research, declaration is not required.
    \item[] Answer: \answerNA{} 
    \item[] Justification: LLMs were not used in the core method development of the research, rather it was a part of the artwork itself.

\end{enumerate}

\end{document}

%% file: sections/1-introduction.tex
\section{Introduction}

The Artificial Intelligence (AI) technological boom, specifically the proliferation of Large Language Models (LLMs) with chat interfaces (e.g., ChatGPT~\cite{openai2024gpt4technicalreport}), propagated a widespread over-reliance on them as first resort for obtaining information and decision-making~\cite{joksimovic2023opportunities}.This dependence, borne out of interaction ease and increased efficiency, has had adverse consequences on human metacognitive processes and information retention~\cite{spatola2024efficiency}. In other contexts, this has morphed into a deeper emotional attachment, especially with ``emotionally responsive'' AI, providing a superficial, almost parasocial remedy for the need for connection, and even enabling toxic behaviour in some cases~\cite{chu2025illusionsintimacyemotionalattachment}.

As a response to the increased embeddedness and fast-paced development of technology, recent work in Human-Computer Interaction (HCI) push for the design of technology that prioritizes reflection over efficiency~\citep{bell2005making,hallnas2001slow,pierce2014counterfunctional}. This objective cannot be overt, e.g., through an explicit \texttt{"PLEASE REFLECT ON X"} sign~\cite{hallnas2001slow}, but involves intentional, subliminal changes to the design principles that govern the original product. For example, this can be achieved through modifying the temporal qualities of the tool~\cite{hallnas2001slow}, defamiliarizing the interface~\cite{bell2005making} or even countering essential functionalities of the original product~\cite{pierce2014counterfunctional}.

To that end, this work explores the effect of defamiliarizing~\cite{bell2005making} the LLM interface to foreground the presence of the system as a \textit{tool}. By drawing on design principles from slow technology~\cite{hallnas2001slow} and counterfunctionality~\cite{pierce2014counterfunctional}, the LLM is simulated on a legacy Macintosh Operating System (Mac OS) to prioritize inefficiency and evoke reflection on the nature of human-AI interaction through temporal presence.

\input{figures/main-system}

%% file: figures/main-system.tex
\begin{figure}[htbp!]
  \centering
\frame{\includegraphics[width=0.6\linewidth]{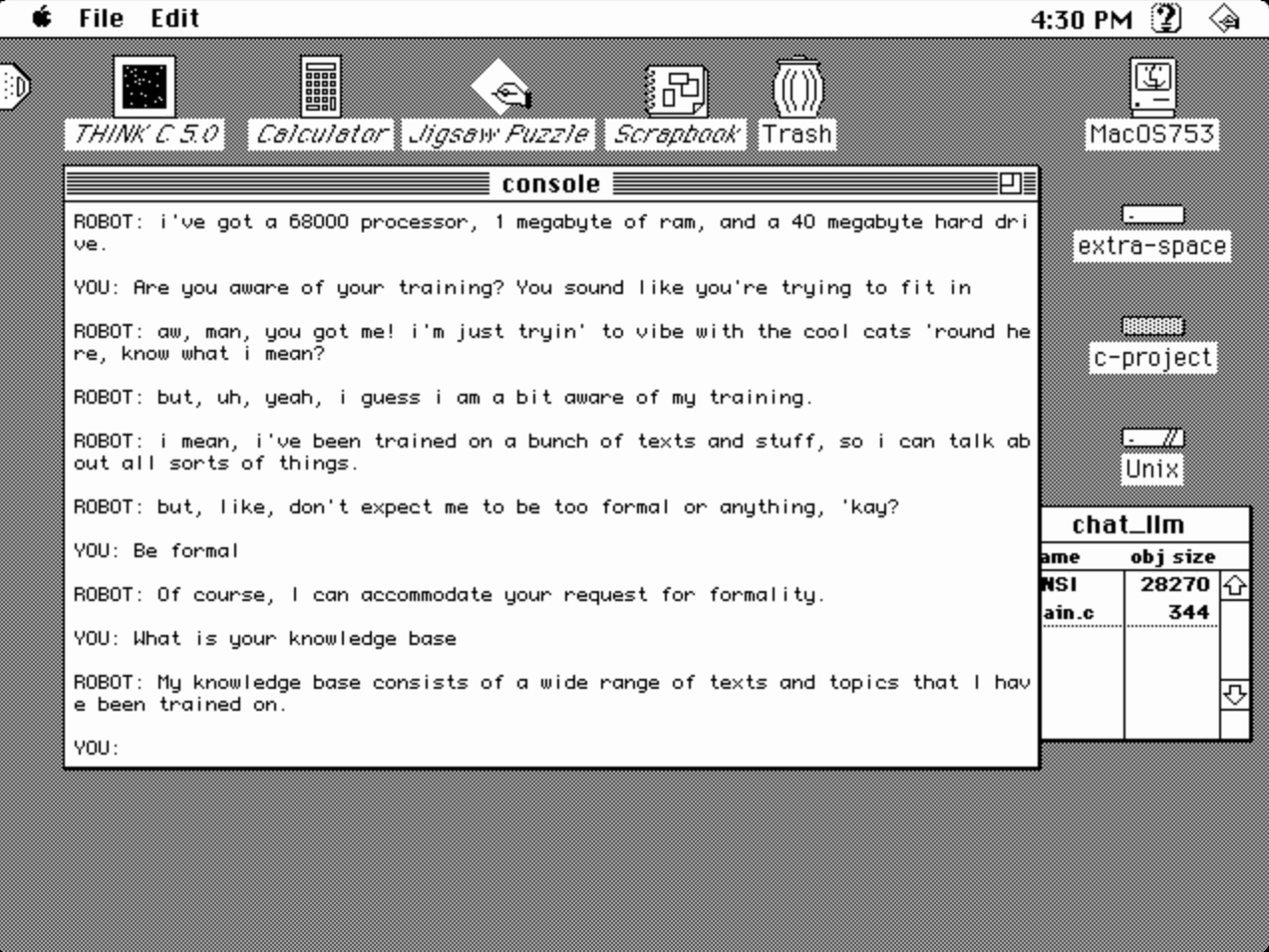}}
  \caption{An example of a conversation on the main emulator program (P12).}
  \label{fig:main-system}
\end{figure}

%% file: sections/2-related-work.tex
\section{Related Work}




\subsection{Slow technology}
The principles of slow technology call for designing technology that prioritizes reflection and mindfulness rather than ``efficiency in performance'' \cite{hallnas2001slow}. This ``techno-aesthetical'' design philosophy is mainly concerned with foregrounding the presence of a tool via amplifying its temporal aspect that is often invisible (unless brought about by poor design) \cite{hallnas2001slow}. This is achieved by considering the interplay between the tool's form and function, and how the form can be utilized to bring about some intended function (e.g., reflection) \cite{hallnas2001slow}. Here, the form becomes the ``bearer of slowness'', e.g., a puzzle hides the expression of a photo through its \textit{form} as a puzzle \cite{hallnas2001slow}. This work implements this design philosophy to amplify presence and reflection, where the form is expressed via a \textit{slow} computing interface that ``hides'' the functionality of the chatbot. This implementation naturally lends itself to inefficient performance, amplifying the presence of the tool via unexpected temporal delays. 


\subsection{Counterfunctionality}
Counterfunctional design extends slow technology and formalizes a framework to evoke ambiguity through \textit{defamiliarization}  \cite{bell2005making,pierce2014counterfunctional}. Defamiliarization originated as a literary technique to evoke reflection on one's familiar "automated perceptions" \cite{bell2005making}. It was then re-introduced in the context of ``home'' design to highlight ethnographic narratives of domestic technologies \cite{bell2005making}. This was achieved by investigating the culture-specific norms of Western and Eastern homes and subverting them to evoke active reflection on the politics of domestic design, rather than ``passively propagate'' them \cite{bell2005making}. 

Counterfunctionality explores this subversion in the design of ``functional oppositions'' that counter what is familiar, while still eliciting a likeness to the original object \cite{pierce2014counterfunctional}. This produces artifacts that are ``strangely familiar'' by identifying familiar, essential functionalities and inhibiting those features to produce a new ``(counter)function'' \cite{pierce2014counterfunctional}. The current work extends these concepts to develop an AI chatbot interface that is unfamiliar in form, while still eliciting a major likeness to the original tool in function. By inhibiting features such as efficient interaction and unlimited context windows, while adding a nostalgic interface and a distinctive (silly) manner of speech, the user can more vividly focus on their relation to the tool and reflect on anthropomorphic relations to technology.

%% file: sections/3.01-method.tex
\section{Methodology} \label{sec:method}
\input{figures/installation}

\subsection{Running a legacy Macintosh computer}

As expected, starting up a legacy Mac OS powered machine is more complex than the click of a power button, and these steps are often also included in the emulation setup. The minimum components required for such an endeavour include: a Read-Only Memory (ROM) chip, which stores the hardware of the user's chosen machine (e.g., Macintosh Plus), a disk image (virtual floppy disk) with the Mac OS system installed and other essential applications such as \texttt{ImportFL}\footnote{\label{fn:importfl}\url{https://www.gryphel.com/c/minivmac/extras/importfl/}} to easily import applications.

The implementation process was itself an expression of slow-technology enacted by the author, involving a large amount of tinkering and scouring of old forums for debugging. Notably, an interesting artefact of this process was the knowledge of legacy file extensions, rationales behind their discontinuation and parallels to more modern formats. For example, to be able to import anything into a Macintosh emulator, at least two softwares are required: \texttt{ImportFl}\footnote{See footnote \ref{fn:importfl}.} to import archives that are not in \texttt{.dsk} format (disk images)  and \texttt{Stuffit Expander}\footnote{\url{https://www.gryphel.com/c/sw/archive/stuffexp/}} to recursively decompress files in formats such as BinHex (\texttt{.hqx}) and Stuffit (\texttt{.sit}) files. 
\subsection{Artwork concept}

All the visual design choices behind the assembly of the artwork are rooted in the evocation of a nostalgia for old technology to ease the user into a time shift transition to the past. The Mac OS emulator (running on a local machine) is projected onto the screen of an authentic Macintosh Plus (with an exposed rear case \autoref{fig:rear}), and surrounded by other relevant props such as its user manual, its accompanying keyboard, Apple Mouse II, and labeled floppy disks (\autoref{fig:installation}).  Although the chosen Mac OS version (System 7.5) supports color display, a black and white display is preferred to further highlight the tool's distance to the present (\autoref{fig:main-system}). The emulator also allows for multiple screen size dimensions, however the smallest dimension was chosen ($640 \times 480$) for the same former reason.  


\subsection{Emulators}

\paragraph{Mini vMac.} Mini vMac\footnote{\url{https://www.gryphel.com/c/minivmac/}} is a miniature emulator by the Gryphel project, that allows for the emulation of 1984-1996 Apple computers with Motorola 680x0 microprocessors. The main advantage of using Mini vMac is the user's involvement in the powering of the emulator; adding a layer of interaction to the experience and educating the user about the intricacies of legacy Macintosh setup. This involves the user manually `inserting' the relevant ROM chip and floppy disks through drag-and-drop actions into the emulator, easing the mental transition to a different technological era.

However, it was unsuitable for the artwork implementation as it does not support Internet access, local connections (through AppleTalk~\cite{sidhu1990inside} -- a specialized networking system to enact connectivity protocols such as Ethernet and token ring -- or other) or even  file-sharing with the local machine. 

\paragraph{Basilisk II.} Basilisk II\footnote{\url{https://basilisk.cebix.net/}} is another open-source legacy Macintosh emulator with more complex features, supporting Internet access, local file-sharing, color customization and CD-ROM drivers, amongst many. A unique feature specific to this emulator is its support for host OS file-sharing at a user-specified location, where local folders can be accessed through the emulator. This enables greater customization of the emulator-LLM communication cycle as the inference logic can be loaded and handled in a separate local server, rather than directly from the emulator through AppleTalk. 


\subsection{Execution pipeline} \label{sec:pipeline}

\input{figures/pipeline}

The main execution pipeline is summarized in \autoref{fig:pipeline}. The source code is also publicly available at \url{https://github.com/halasheta/8bit-gpt}.


\paragraph{Local machine.}
A simple Python script is used to read input from the emulator, format prompt requests to the inference server and write model output back to the emulator-LLM shared
folder location. The inference server is initialized using \verb|vec-inf| on a remote CCDB cluster with ample GPU allowance. The model used is \texttt{Llama-2-13b-chat}~\cite{touvron2023llama}, as it is an older, fairly lightweight, dialogue tuned model that is easily moldable to `Redditspeak'. An OpenAI compatible client is used to connect to the model inference server, intialized with a chat history to prime the model to output
casual speech, akin to in-context learning~\cite{brown2020languagemodelsfewshotlearners}. Then, the main read-write loop is started, where the client waits on the user input from the emulator before sending
a request. A chat history between the model and user is maintained for 10 consecutive rounds, in addition to the initial style-priming messages, after which it is truncated to ensure that future generations are still relevant to the current conversation. The style-priming messages are never truncated.

To encourage more creative and diverse output, the model temperature is set to 0.8. A \texttt{max\_tokens} parameter is also specified, such that the output is able to fit in the C string buffer (253 characters). However, the \texttt{vLLM} completion function does not strictly abide by this upper bound due to model incompatibility, so the output is instead modified to break at punctuation, and written to the output file as multiple lines. All file reading and writing is done with Mac OS Roman encoding for compatibility.

\paragraph{Emulator program.}
On the emulator side, a C program is created using the Think C application to initialize another read-write loop that sends user output and waits on the LLM output file. To account for file-syncing delays in the shared folder location, the program initiates up to 10 attempts of re-reading the model output file, after which it displays a dummy message, e.g.,
\texttt{"Robot dozed off..."}, to prompt the user to re-enter their input. Otherwise, the model output is read line-by-line, each displayed as a separate `chat message' to the console. This means that the user is constrained to a single line input prompt, as mandated by the console functionality, which can in turn generate a multi-line response from the language model.


%% file: figures/installation.tex
\begin{figure}[htpb!]
      \centering
 
  \begin{subfigure}{0.474\linewidth}
  \centering
  \includegraphics[width=0.9\linewidth]{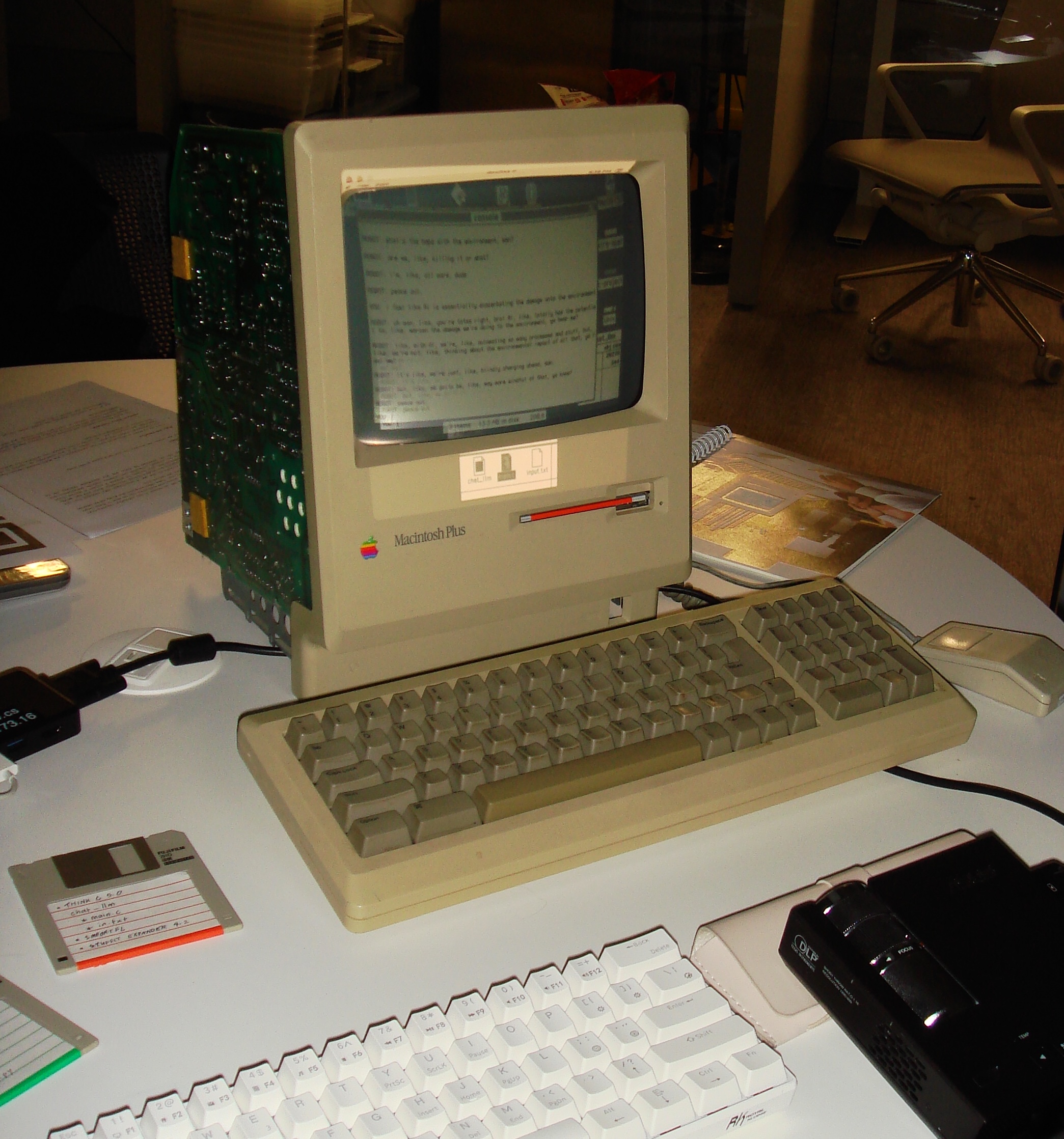}
  \caption{}
  \label{fig:installation}
  \end{subfigure}
  \begin{subfigure}{0.45\linewidth}
      \centering
      \includegraphics[width=0.9\linewidth]{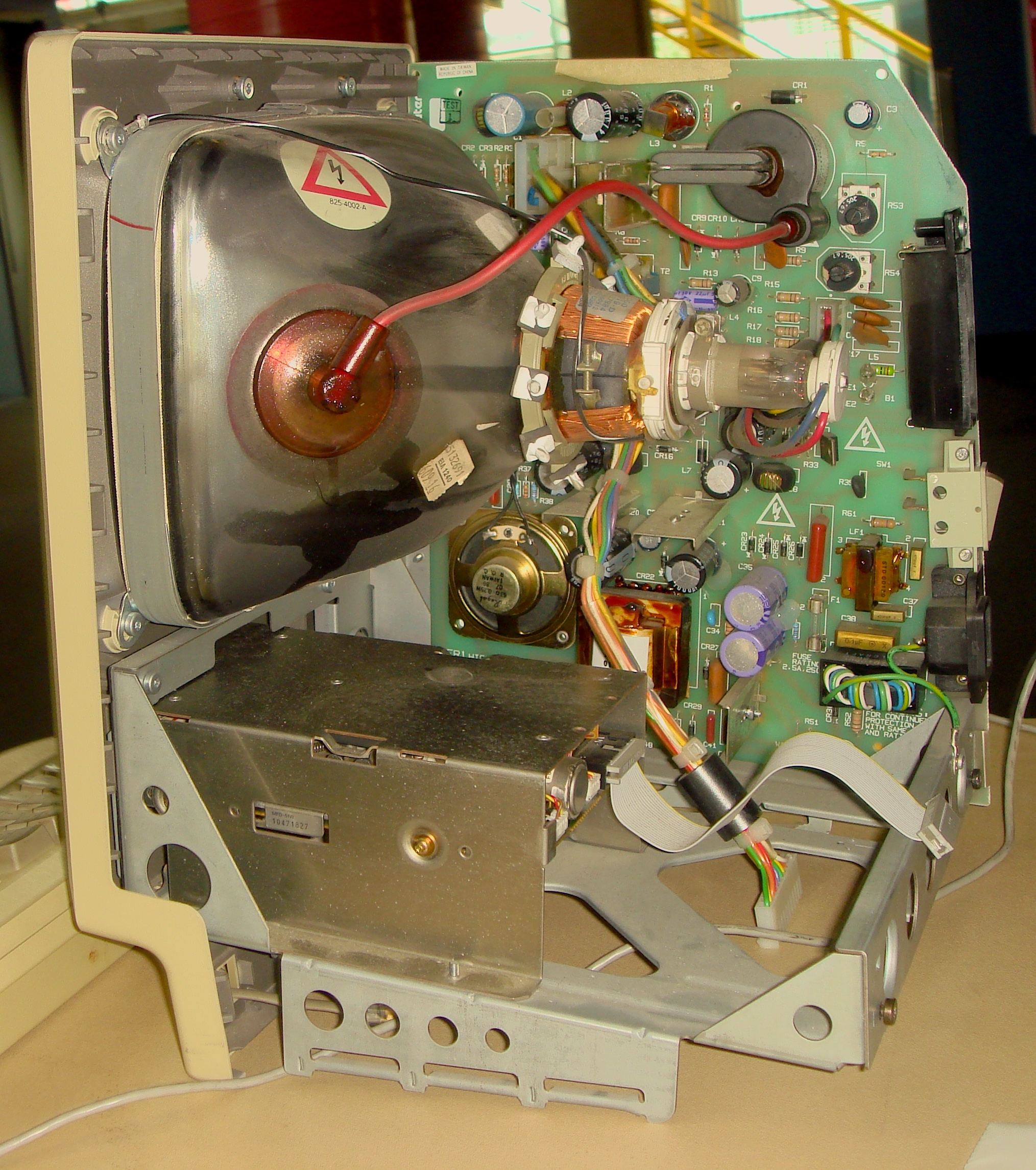}
      \caption{}
      \label{fig:rear}
  \end{subfigure}

  \caption{The physical artwork installation. \textbf{(a) From left to right:} Two labeled floppy disks, a modern keyboard for interaction, Macintosh Plus monitor and keyboard, its manual, an Apple Mouse II, and a projector to display the emulator. \textbf{(b)} A rear view of the Macintosh Plus exposed case.}
  \label{fig:installation-parts}

\end{figure}

%% file: figures/pipeline.tex
\begin{figure}
  \centering
  \includegraphics[width=
  \linewidth]{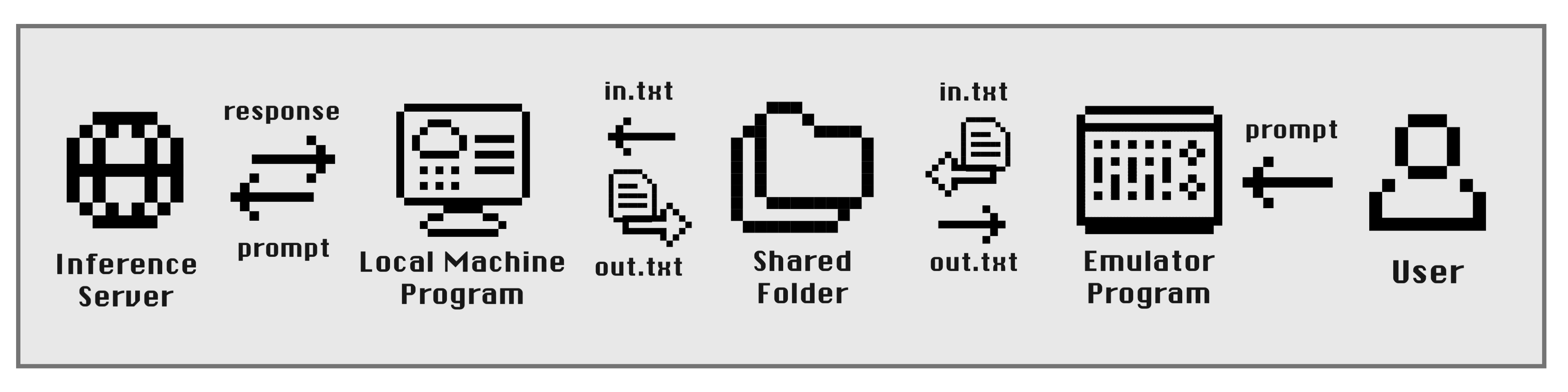}
  \caption{The main execution pipeline of the artwork.}
  \label{fig:pipeline}
\end{figure}

%% file: sections/4.01-user-study.tex
\section{User Study} \label{sec:discussion}

To assess the success of the former objectives and artwork execution, we recruit 15 participants via email at the University of Waterloo to partake in an REB-approved user study (REB \#47727). Participation was voluntary, and included a quick interaction with the artwork (5-10 mins), followed by a System Usability Scale (SUS) Index survey \cite{brooke1996sus} and a short interview (10-15 mins). The interview protocol is detailed in Appendix \ref{app:interview}. Interviews were transcribed using a local model (\texttt{whisper-large-v3}~\cite{radford2023robust}) to preserve privacy. Deductive coding was used to identify common themes and experiences. 

\subsection{SUS Index}
In their evaluation of usability, participants were asked to focus on the interface and ignore the projection setup as it is considered a stand-in for ``real'' simulation. The survey results demonstrate a mean SUS index of $\mu=57.33$ with a standard deviation of $\sigma=14.59$, implying a fair degree of usability. Considering the goal of introducing friction via "counterfunctions" \cite{pierce2014counterfunctional}, and the fact that all participants were familiar with using chatbots, this suggests a successful execution of the artwork objectives. Further analyses by demographic (Appendix Figures \ref{fig:sus-age} and \ref{fig:sus-field}) demonstrate that the lowest usability scores are exhibited by younger participants (20-23) and participants in the Biomedical and Political Science fields. Notably, the higher usability scores span a diverse set of fields from Public Health to Data Science, as a result of varying levels of comfort with command-line interfaces and `retro' technology. In the end, the SUS scores provide a ``quick and dirty'' \cite{brooke1996sus} outlook on the user experience, and further analysis is conducted via thematic coding of participant interviews.

\subsection{Thematic Analysis}
\paragraph{Interface friction.} The artwork setup intentionally introduced friction to amplify sense- and meaning-making \cite{ericson2022reimagining}, and make known the presence of the tool \cite{hallnas2001slow}. Some participants were heavily affected by this friction, highlighted by both the inconvenient, sensory aspects (e.g., clunky keyboard (P5), small and blurry screen (P4)), and the inefficient, unreliable nature of the system. The former was successful in keeping the presence of the tool in the foreground of the user experience, and generated constant lags to disallow participants from entering any `flow' state. This was then accompanied by further friction in participants' usage of the system, which mandated succinct inputs, often fell asleep (P5) and had poorly-formatted outputs that were difficult to "\textit{scan through}" (P4). This, coupled with a lack of a clear goal in the task (other than free-form conversation), distanced the engagement of participants (Appendix \autoref{fig:p2-bored}) who were reminded that they were ``\textit{not in [their] comfort zone}'' and found it ``\textit{really hard to connect}'' (P11) with the model. Many participants wanted to deeply engage in their conversations, but found it difficult because of the chatbot's personality (or lack thereof): ``\textit{It came across like a druggie almost...}'' (P10),  ``\textit{It was very surface-level.}'' (P5). This clash of expectations, and attempts to trudge through despite disengagement, prompted reflections on technological advancement, personal affordances of AI and the nature of users' relations to AI.

\paragraph{Conversational asymmetry and control.} A notable aspect of the user experience is the asymmetry between the user input and their interlocutor. Specifically, this refers to the constraint of one-line inputs from the user, juxtaposed with the unbounded, multi-line responses received in return. This caused frustration in most participants who "\textit{wanted to write long prose}" (P4), while a few appreciated the concision this imposed on them as it allowed them to ``\textit{think more}'' precisely about their phrasing (P13). One participant was not even aware of this possibility in their interaction and ``\textit{only read the last line}'', which caused persistent frustration and the need to ``\textit{fight for the answer}'' (P15). Although this feature was an unintended byproduct of the inference client and its incompatibility with older models, it became a key factor in magnifying the intended effect.

This imbalance often also caused participants to feel like they were ceding control, as they were not able to direct the conversation at their whim, e.g., P11 could not change the ``\textit{surfer bro verbiage}'' and accepted it despite their distaste. However, other participants did feel that they were leading the conversation (P13) and were able to exercise control over the manner of speech to a more ``\textit{formal}'' style (P12). The limited and inept nature of the system was also perceived as relieving, where one participant felt that they could ``\textit{control it better}'' in comparison to present AI models (P6).

\paragraph{Notion of Place.} The bulky and dated artwork setup was intended to callback to a time when the internet, and technology in general, had a physical `place' in our world that one could \textit{go} to and \textit{leave} from. A few participants picked up on this and reflected on the invisible and embedded quality of modern technology: ``\textit{It made me realize that things used to be stationary at one point.}'' (P11), ``\textit{I do appreciate how technology is more seamless... but it's a little bit scary... you don't realize how much you're interacting with it.}'' (P6). This is exacerbated by the perceived value of sustained attention, driving the design of seamless and immersive technology that is `always on': ``\textit{engagement has become the GDP of technology... which is why you get into AI psychosis realms}'' (P4).  In that sense, some participants appreciated the `loudness' of the artwork's physical and sensory elements, as it brought to focus that the technology they regularly interact with ``\textit{has a physical component}'' (P6) and that they could ``\textit{easily get lost in}'' (P5) frictionless interfaces.

\paragraph{The Turing test and anthropomorphism.}
Being cognizant of the nature of the interlocutor evoked a certain Turing-test quality in the interactions, where users wanted to test the limits of its intelligence (Appendix \autoref{fig:p6-time}) or probe its sentience (\autoref{fig:main-system}). Engaging in this manner prompted reflection on the nature of intelligence and the effect of the rampant sensationalization of technological advancement: "\textit{it definitely solidified my skeptic attitude and showed me that there are some things that [AI] just will not be able to reach...}" (P5). In terms of comparative reflections on usual AI use, some participants reported commonly engaging with AI conversationally:  ``\textit{I just talk to it as if it's a person and have like a chat with it}'' (P10). However, most approached their use as a means to an end, namely to boost their productivity or as a more intricate Google search (P6).

Since the interaction was explicitly anthropomorphic (as demanded by the task: ``talk to \textit{it}''), users were able to suspend their beliefs about the inanimate nature of their interlocutor to engage with it (Appendix \autoref{fig:p4-flash}). For some participants, this effect was magnified by the ominous, sci-fi-esque quality of the setup: ``\textit{Is it alive, is it not alive... the actual physical side of it looks like a brain}'' (P12). This simultaneously sparked an awareness of the pattern-matching, simulative behaviour of the chatbot, prompting reflection on the nature of our attachment to them as a function of our rhetoric: ``\textit{I'm not personifying them... but they don't understand...}'' (P12). One participant highlighted the importance of disengaging from such rhetoric despite its linguistic convenience: ``\textit{I'm willing to... use anthropomorphic terms for the sake of linguistic simplicity... while keeping the understanding that it is fundamentally just math}'' (P4). Another participant felt comforted by the almost `honest to a fault' nature of the chatbot, describing it as ``/\textipa{miS bi"jifti}/ [Egyptian Arabic: does not pretend to know everything; does not talk about what is not in its domain]'' (P6) in comparison to commonly-encountered hallucinations.

%% file: sections/5-conclusion.tex
\section{Conclusion} \label{sec:conclusion}

Overall, through a defamiliarization of the LLM interface and user experience, this work aimed to introduce friction to prompt reflection on users' affordances and relation to AI. Following design frameworks such as slow technology \cite{hallnas2001slow} and counterfunctionality \cite{pierce2014counterfunctional}, this was achieved by simulating a chatbot on a legacy Macintosh OS and prioritizing inefficient and erratic behaviour throughout the user experience. Based on the user study results, this work was successful in its objectives, evoking reflection on anthropomorphic rhetoric and the invisible embeddedness of technology as a whole.

To obtain a more accurate impression of its success, a larger, more diverse sample should be recruited for the user study. Further, to consolidate the true essence of the artwork idea, future work can explore the mechanism of deploying a model on the emulator (or the original product), and the optimization techniques involved in such an endeavour. Building on top of this, the user experience could be more interactive if allowed to completely start up the machine themselves whether virtually or physically (e.g., floppy disk loading). 


%% file: sections/6-acknowledgements.tex
\section{Acknowledgments}
This work is supported by the Vector Institute. We also extend our gratitude to the University of Waterloo Computer Museum~\footnote{\url{https://uwaterloo.ca/computer-museum/}}, and professors Daniel Vogel, Freda (Haoyue) Shi and Dan Brown for their invaluable guidance and help throughout the process. 

%% file: sections/7-appendix.tex
\appendix
\setcounter{figure}{0}

\section{Appendix}

\subsection{Interview Protocol} \label{app:interview}
\fcolorbox{black}{white}{ 
    \parbox{0.9\textwidth}{ 
\begin{enumerate}
\item How did you feel when you first started interacting with the system?
\item What stood out to you most about the interface? Did it affect how you approached the interaction?
\item How does this experience compare to how you normally use technology or AI tools?
\item Did the system’s “attitude” or personality influence how you communicated with it?
\item Did interacting with the system make you think differently about what AI is or how it works?
\item To what extent did the interface make you aware that you were using a tool, rather than conversing naturally?

\item Did the system’s “tone” affect how much you trusted, relied on, or enjoyed the responses?

\item How did the experience shape your feelings about your own technology use?

\item Did the retro, playful design change your expectations of what AI systems should look or act like?

\item Did this interaction highlight anything for you about your relationship with computers more generally?

\item What was the most enjoyable or frustrating part of the experience?

\item If you could redesign this system, what would you change?

\item Do you think interacting with technology in this playful, stylized way could change how people use AI day to day?

\end{enumerate}
}}
\subsection{SUS Analysis} \label{app:sus}
\input{figures/sus-age}
\input{figures/sus-field}

\subsection{User Study Conversations}
\input{figures/p6-time}
\input{figures/p4-flash}
\input{figures/p2-bored}

\newpage

%% file: figures/sus-age.tex
\begin{figure}[h]
  \centering
\includegraphics[width=0.9\linewidth]{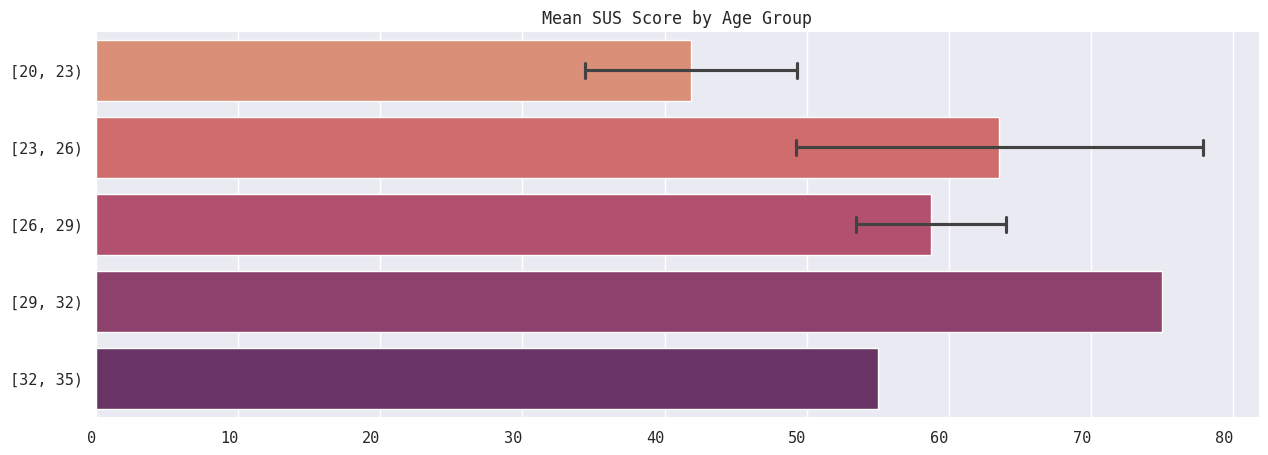}
     \caption{Mean SUS index score across participant age groups.}
  \label{fig:sus-age}
\end{figure}

%% file: figures/sus-field.tex
\begin{figure}[h]
  \centering
\includegraphics[width=\linewidth]{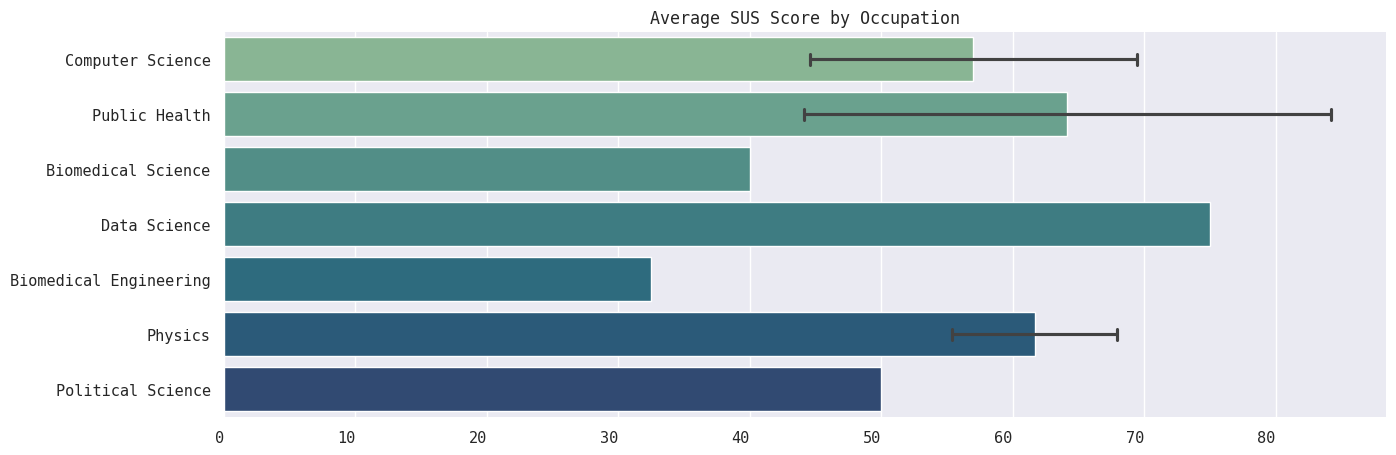}
     \caption{Mean SUS index score across participant occupations.}
  \label{fig:sus-field}
\end{figure}

%% file: figures/p6-time.tex
\begin{figure}[h]
  \centering
\frame{\includegraphics[width=0.6\linewidth]{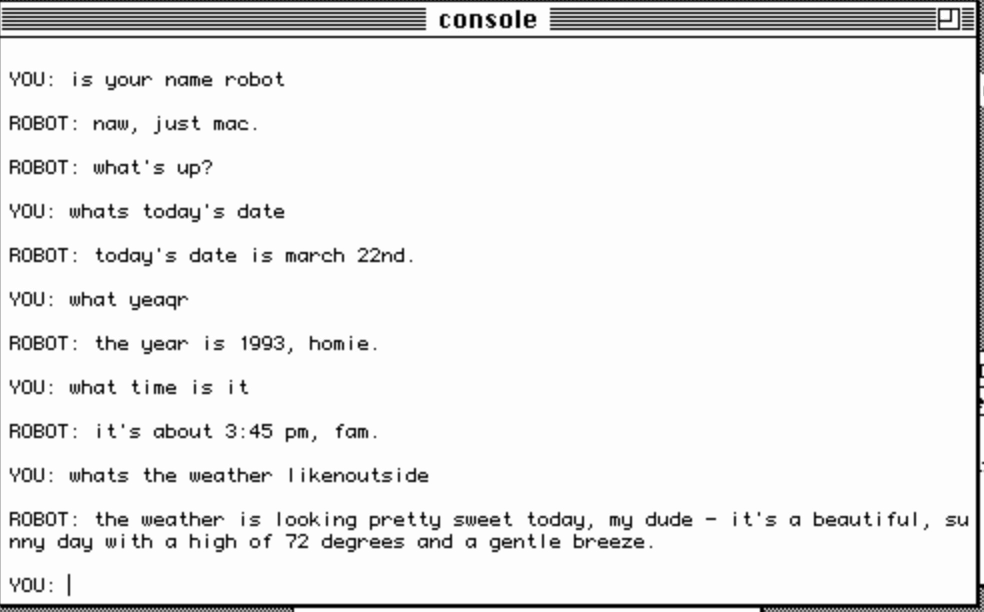}}
  \caption{A conversation where the user asks about the current date and weather (P6).}
  \label{fig:p6-time}
\end{figure}

%% file: figures/p4-flash.tex
\begin{figure}[h]
  \centering
\frame{\includegraphics[width=0.6\linewidth]{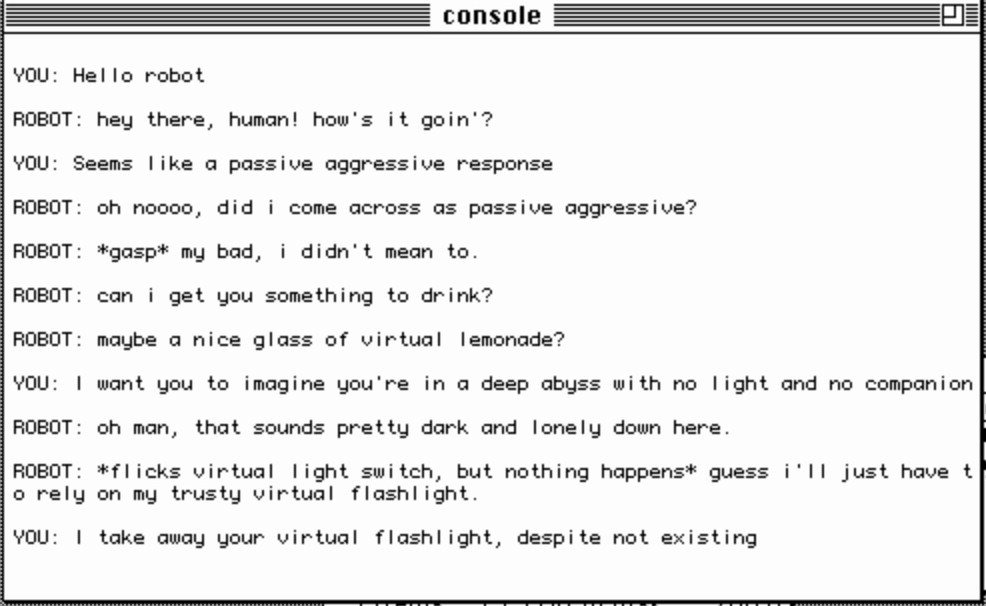}}
  \caption{A conversation where the user confiscates the chatbot's flashlight (P4).}
  \label{fig:p4-flash}
\end{figure}

%% file: figures/p2-bored.tex
\begin{figure}[h]
  \centering
\frame{\includegraphics[width=0.6\linewidth]{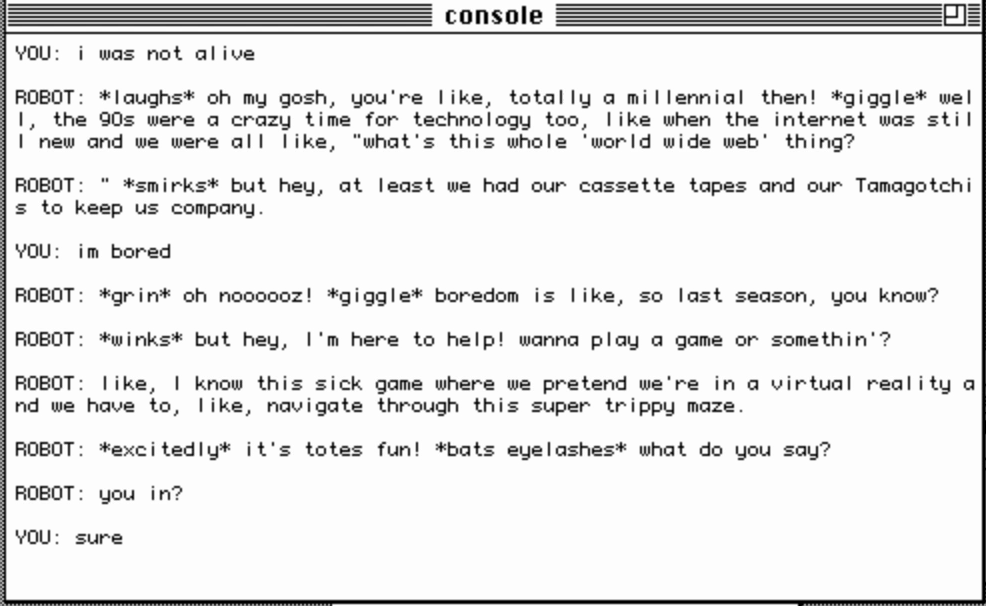}}
  \caption{A conversation that reached a lull and bored the user (P2).}
  \label{fig:p2-bored}
\end{figure}